\documentclass[aps,prl,reprint,longbibliography]{revtex4-1}
\usepackage{amsmath}
\usepackage{amssymb}
\usepackage{graphicx}
\usepackage{hyperref}
\usepackage{url}
\usepackage{color,xcolor}
\usepackage{ulem}

\begin{document}
\title{Appearance and disappearance of ferromagnetism in ultra-thin LaMnO$_3$ on SrTiO$_3$ substrate: a viewpoint from first-principles}
\author{Ming An}
\author{Yakui Weng}
\author{Huimin Zhang}
\author{Jun-Jie Zhang}
\author{Yang Zhang}
\author{Shuai Dong}
\email{Correspondence: sdong@seu.edu.cn}
\affiliation{School of Physics, Southeast University, Nanjing 211189, China}
\date{\today}

\begin{abstract}
The intrinsic magnetic state (ferromagnetic or antiferromagnetic) of ultra-thin LaMnO$_3$ films on the mostly used SrTiO$_3$ substrate is a long-existing question under debate. Either strain effect or non-stoichiometry was argued to be responsible for the experimental ferromagnetism. In a recent experiment [Science \textbf{349}, 716 (2015)], one more mechanism, namely the self-doping due to polar discontinuity, was argued to be the driving force of ferromagnetism beyond the critical thickness. Here systematic first-principles calculations have been performed to check these mechanisms in ultra-thin LaMnO$_3$ films as well as superlattices. Starting from the very precise descriptions of both LaMnO$_3$ and SrTiO$_3$, it is found that the compressive strain is the dominant force for the appearance of ferromagnetism, while the open surface with oxygen vacancies leads to the suppression of ferromagnetism. Within LaMnO$_3$ layers, the charge reconstructions involve many competitive factors and certainly go beyond the intuitive polar catastrophe model established for LaAlO$_3$/SrTiO$_3$ heterostructures. Our study not only explains the long-term puzzle regarding the magnetism of ultra-thin LaMnO$_3$ films, but also shed light on how to overcome the notorious magnetic dead layer in ultra-thin manganites.
\end{abstract}
\maketitle

\section{I. Introduction}
Perovskite oxides in the form of $AB$O$_3$ are important materials, which cover a wide range of exotic physical properties including unconventional superconductivity, colossal magnetoresistance, multiferroicity, and so on \cite{Imada:Rmp,Dagotto:Sci,Dong:Ap}. Recent technical advances in film fabrications have enabled the atomic-level construction of various perovskite oxides and their heterostructures \cite{Schlom:Armr,Martin:Mse,Bibes:Ap,Huang:Mplb,Zubko:Arcmp,Chakhalian:Nm}, which lead to new emergent physics at the interface/surface and shed light on potential electronic devices based on oxides \cite{Dagotto:Sci07,Mannhart:Sci,Hammerl:Sci,Takagi:Sci,Hwang:Nm}. By reducing the thickness and dimension, perovskites can exhibit distinct physical properties from the corresponding bulks, which are physical interesting and important for applications \cite{Dagotto:Bok,Tokura:Rpp}.

LaMnO$_3$ (LMO), as the parent compound of colossal magnetoresistance manganites \cite{Wollan:Pr}, is one of the mostly studied perovskite oxides with abundant physical properties \cite{Carvajal:Prb,Dagotto:Prp,Salamon:Rmp}, which is also widely used as a building block in oxide heterostructures  \cite{May:Nm,Bhattacharya:Prl,Dong:Prb08.3,Aruta:Prb,Nanda:Prb,Nanda:Prb09,Yamada:Apl,Garcia:Am,Zhang:Prb12,Dong:Prb12,Choi:Prb,Gibert:Nm,Dong:Prb13,Zhai:Nc}. The bulk of LMO was reported to be A-type antiferromagnetic (A-AFM) at low temperatures, namely spins are parallel in the $ab$ plane but antiparallel between nearest neighbors along the $c$-axis \cite{Wollan:Pr}. However, this A-AFM is quite fragile against tiny non-stoichiometry, and thus sometimes obvious ferromagnetic (FM) signal is observed even for single crystalline samples \cite{Ritter:Prb}. In many experiments of LMO films on the mostly used SrTiO$_3$ (STO) substrate, FM insulating behavior has been observed \cite{Bhattacharya:Prl,Gibert:Nm,Zhai:Nc,Choi:Jpd,Orgiani:Apl}, which has been under debate for a long time. Both the strain effect and non-stoichiometry have been proposed to explain this A-AFM to FM transition \cite{Dong:Prb08.3,Hou:Prb,Bhattacharya:Prl,Gibert:Nm,Zhai:Nc,Choi:Jpd,Orgiani:Apl}.

Theoretically, many attempts suggest that the FM phase is intrinsic for STO-strained LMO \cite{Dong:Prb08.3,Hou:Prb,Lee:Prb13,Nanda:Jmmm}. For example, double-exchange model calculations suggested the possible FM orbital-ordered (OO) phase for cubic LMO \cite{Hotta:Prb03,Dong:Prb08.3}. Based on density functional theory (DFT) calculations, Hou \textit{et al.} proposed a new OO phase driven by strain as the FM insulating LMO \cite{Hou:Prb}. Another recent DFT calculation by Lee \textit{et al.} also claimed a FM phase for strained LMO although it was metallic \cite{Lee:Prb13}.

Recently, Wang \textit{et al.} synthesized high-quality epitaxial ultra-thin LMO films on TiO$_2$ terminated [001]-oriented STO substrate and observed an atomically sharp transition from a no-magnetization phase to FM phase when the thickness of LMO reached five unit cells (u.c.) \cite{Wang:Sci349}. This thickness dependent magnetic transition was argued to be the result of charge reconstruction induced by polar discontinuity \cite{Wang:Sci349}, since there was also a similar critical thickness in the polar catastrophe model for the (001)-orientated LaAlO$_3$/STO heterostructures \cite{Nakagawa:Nm,Delugas:Prl,Chen:Am}. This scenario, in which the FM phase is born from the intrinsic non-FM background due to the self-doping (i.e. electrons transfer from surface to interface), is in contradiction with aforementioned theoretical results \cite{Dong:Prb08.3,Hou:Prb,Lee:Prb13}. However, if these theoretical calculations are correct, i.e. strained LMO film is intrinsically FM, it is a puzzle for the existence of a critical thickness in Wang \textit{et al.}'s experiment, below which the FM signal disappears.

In fact, the disappearance of FM magnetization in ultra-thin FM perovskite films, like doped manganite La$_{1-x}$Sr$_x$MnO$_3$ (LSMO) and SrRuO$_3$, is also a long-time puzzle with many debates \cite{Sun:Apl,Borges:Jap,Kavich:Prb,Huijben:Prb,Tebano:Prl,Kourkoutis:Pnas,Gray:Prb,Song:SIA,Peng:Apl,Xia:Prb,Chang:Prl}. This phenomenon is called magnetic ``dead layers". The solution of magnetic ``dead layers" is crucially important for the pursuit of magnetic oxide electronics. Therefore, as the first step, it is of particular importance to understand the physical mechanism of the thickness dependent magnetism in ultra-thin magnetic perovksite films.

In this work, the structural and magnetic properties of (001)-oriented vacuum/(LMO)$_n$/(STO)$_2$ heterostructures will be studied using DFT calculation, in order to reveal the mechanism behind the thickness dependent magnetic transition. It should be noted although there have been many DFT studies on strained LMO, some of these calculations have not really put STO layer and vaccum into considerations \cite{Nanda:Jmmm,Hou:Prb,Lee:Prb13}, or those previous DFT methods would be failed to correctly describe the strain effect in LMO/STO heterostructures \cite{Jilili:Sr,Liu:Jap}, as clarified in the following section. In addition, the (LMO)$_n$/(STO)$_m$ superlattices without surface will also be calculated. The comparison between these two series of LMO/STO heterostructures can highlight the vital role of surface to the magnetic phase transition in ultra-thin LMO.

\section{II. Model \& Method}
LMO is a Mott insulator with the orthorhombic $Pbnm$ structure \cite{Carvajal:Prb}. STO is a band insulator with cubic perovskite lattice, whose lattice constant ($a_{\rm STO}$) is about $3.905$ {\AA}. The STO substrate provides a compressive strain ($\sim-2\%$) for the LMO film.

In the following, two series of LMO/STO heterostructures will be studied. The first series are superlattices constructed as (LMO)$_n$/(STO)$_m$ ($n$=$1$, $2$, $3$, $4$, $5$; $m$=$1$, $2$, $3$), in which atoms are periodically stacked along the [001] direction without surface. The second series are constructed as (LMO)$_n$/(STO)$_2$ ($n$=$1$, $2$, $3$, $4$, $5$) with open surface (simulated by inserting a $15$ {\AA} vacuum layer). For the open surface heterostructures, the TiO$_2$ termination at interface is adopted, considering the experimental practice \cite{Wang:Sci349}.

To mimic the epitaxial stain from substrate STO, the in-plane lattice constants are fixed as $a=b=\sqrt{2}a_{\rm STO}=5.5225$ {\AA} for both superlattices and open-surface heterostructures. The out-of-plane lattice constant $c$ and the ionic coordinates are fully relaxed to reach the equilibrium state. For the open surface ones, the bottom atom layer of STO substrate (i.e. SrO) is fixed during the relaxation to mimic a very thick substrate. Thicker STO layers have also been checked, which do not change the conclusion (see Supplemental Materials for more details \cite{Supp}). Therefore, two STO layers with fixed bottom are enough to simulate the substrate, which will be adopted in the following calculations.

\begin{figure}
\includegraphics[width=0.48\textwidth]{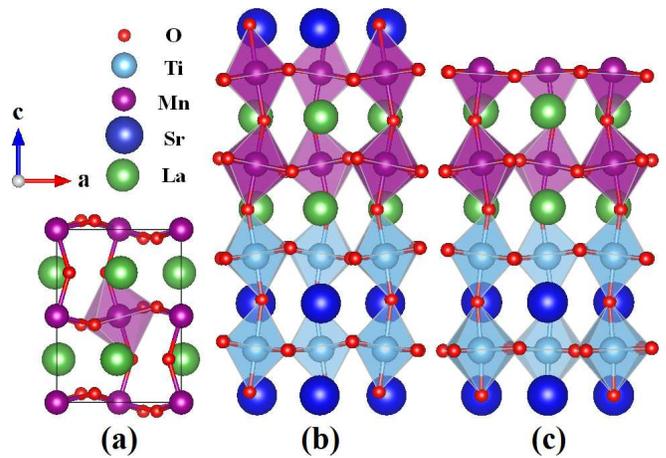}
\caption{Crystalline structures. (a) LMO bulk. (b) (LMO)$_2$/(STO)$_2$ superlattice. (c) (LMO)$_2$/(STO)$_2$ heterostructure with open surface. Here TiO$_2$ layer is the interfacial termination and thus MnO$_2$ is the surface termination.}
\label{fig1}
\end{figure}

First-principles calculations based on DFT are performed using the generalized gradient approximation (GGA) \cite{Perdew:Prl} with Perdew-Burke-Ernzerhof functional modified for solids (PBEsol) parametrization as implemented in Vienna \textit{ab initio} simulation package (VASP) \cite{Perdew:Prb,Blochl:Prb2,Kresse:Prb96,Kresse:Prb99}. The on-site Coulomb corrections $U$ and $J$ are applied to Ti's and Mn's $3d$ orbitals, whose values has been systematically tested to reproduce the experimental lattice parameters and magnetic properties of bulk STO and LMO (see Supplemental Materials for more details \cite{Supp}). The $U$ and $J$ on La's $4f$ electrons have also been tested, which do not affect the physical results of LMO, and thus will be neglected in the following calculations. The plane-wave cutoff energy is set to be $500$ eV. The $\Gamma$-centered $k$-point meshes are set to $7\times7\times5$ and $11\times11\times11$ for LMO and STO bulks respectively. While $5\times5\times2$ mesh and $5\times5\times1$ mesh are accordingly adapted for LMO/STO heterostructures. The atomic positions are optimized iteratively until the Hellmann-Feynman force on each atom is less than $0.01$ eV/{\AA} during structural relaxation. The dipole moment correction is also tested for heterostructures with open surfaces. However, this item only slightly affects the energy of our system and does not change any physical conclusions.

\section{III. Results}
\subsection{III.A. Tests of LMO and STO}
Before the simulation of heterostructures, it is essential to check the physical properties of LMO and STO bulks, which is not a trivial task \cite{Sawada:Prb56,Hashimoto:Prb,Hou:Prb,Lee:Prb,Mellan:Prb,Rivero:Prl}.

\begin{figure}
\includegraphics[width=0.48\textwidth]{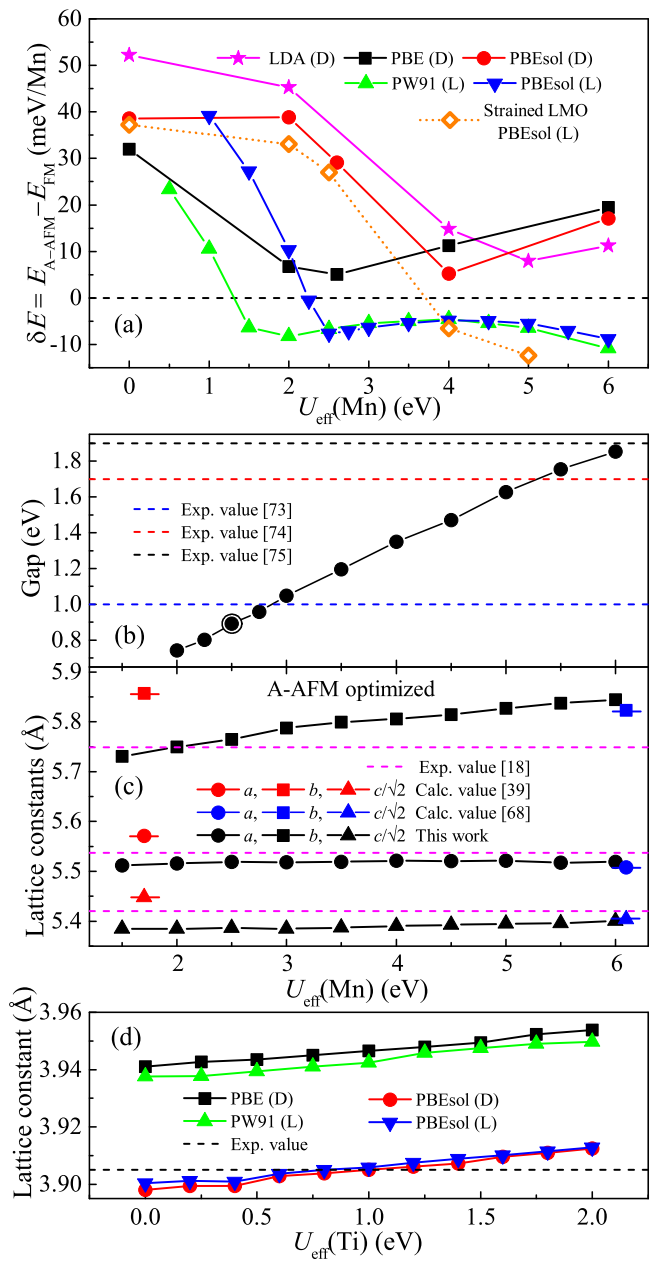}
\caption{(a) The energy difference between A-AFM and FM states for LMO bulk. Different pseudopotentials and adding-$U$ methods have been compared. L: Liechtenstein's method; D: Dudarev's method. Other magnetic states (e.g. G-type AFM and C-type AFM) are also tested, which are much higher in energy and thus not shown here. The strained LMO case is also shown as the orange dashed curve. (b) The calculated band gap of bulk LMO with respect to different $U_{\rm eff}$(Mn). For comparison, several experimental values are also presented as dashed lines. (c) The optimized lattice constants of LMO for the A-AFM ground state. The experimental values \cite{Carvajal:Prb,Mellan:Prb} and previous DFT values \cite{Lee:Prb} are shown for comparison. (d) The optimized lattice constant of cubic STO. Here $\frac{J(\rm Mn)}{U(\rm Mn)}$ is fixed as $0.286$, while $\frac{J(\rm Ti)}{U(\rm Ti)}$ is fixed as $0.273$. More discussions on the choice of $\frac{J}{U}$ can be found in Supplemental Materials \cite{Supp}.}
\label{fig2}
\end{figure}

First, it is well known that the magnetic ground state of LMO is difficult to be captured in DFT calculations \cite{Sawada:Prb56,Hashimoto:Prb,Wang:Sci349}. If the frequently used Dudarev implementation \cite{Dudarev:Prb} is adopted to apply $U_{\rm eff}$(=$U$-$J$), the ground state is always the FM after the structure relaxation, despite the value of $U_{\rm eff}$ and the choice of pseudo-potentials (PBE or PBEsol), as shown in Fig.~\ref{fig2}(a). This bias to FM was repeatedly reported by previous literature \cite{Sawada:Prb56,Hashimoto:Prb,Wang:Sci349}, and sometimes artificial energy compensation was used to move the phase boundary \cite{Wang:Sci349}. However, such artificial operation makes it not convincing to predict new physics or understand the correct physical mechanism.

Second, there are other methods to obtain the A-AFM for LMO. For example, Hou \textit{et al.} \cite{Hou:Prb} used the PW91 pseudo-potentials and rotationally invariant LSDA+$U$ introduced by Liechtenstein \textit{et al.} \cite{Liechtenstein:Prb}. And Lee \textit{et al.} used PBE pseudo-potentials and also the Liechtenstein method to apply $U$ and $J$ \cite{Lee:Prb13}. Both these calculations can lead to the A-AFM ground state. However, their choices will lead to large deviation for STO lattice constant, as shown in Fig.~\ref{fig2}(d). Furthermore, in their calculations, the lattice constants of LMO did not precisely match the experimental ones either (see Fig.~\ref{fig2}(c)). Thus, there were uncertainties regarding the strain between STO and LMO layers in these calculations. These subtle deviations may obstruct the correct understanding of magnetism in LMO/STO heterostructures, since LMO itself is just staying at the edge of phase boundary.

Third, the new developed PBEsol potentials can give much improved precision to describe the crystalline structure for bulks \cite{Perdew:Prl08}. By using the Liechtenstein method to apply apply $U$/$J$ and the PBEsol potentials, our structural optimization can properly reproduce the A-AFM ground state, and obtain the very precise structural information for both LMO and STO, in proper $U$ ranges, as shown in Fig.~\ref{fig2}. For example, when $U$(Mn)=$3.5$ eV and $J$(Mn)=$1$ eV, our calculation gives A-AFM as the ground state, whose energy is lower than the FM state for $7.8$ meV/Mn. The deviations of lattice constants from low temperature experimental values \cite{Carvajal:Prb} are within $-0.3\%$, $0.4\%$, $-0.6\%$ for $a$, $b$, $c$, respectively. And when $U$(Ti)=$1.2$ eV and $J$(Ti)=$0.4$ eV, the lattice constant of calculated STO is just $3.905$ {\AA}. These results pave the solid bases for following calculations on LMO/STO heterostructures. Only the precise structures can correctly describe the strain within LMO/STO heterostructures. In this sense, our results on LMO and STO provide a reliable starting point to reveal the physical mechanism of magnetic transition. In the following, these $U$'s and $J$'s will be adopted by default.

In addition, the estimated local magnetic moment is about $3.6$ $\mu_{\rm B}$ per Mn, which is close to the expected high spin value \cite{Dagotto:Prp,Tokura:Rpp}.

The calculated band gap of LMO is shown in Fig.~\ref{fig2}(b). Several experimental values from different groups are also presented for comparison \cite{Arima:Prb,Saitoh:Prb,Jung:Prb}. Our chosen $U$(Mn) and $J$(Mn) lead to $\sim0.9$ eV, lower than all experimental values more or less. It is a well known drawback that DFT calculations always underestimate band gaps, especially for correlated electron systems. In Ref.~\cite{Mellan:Prb}, a very large $U$ ($U$=$8$ eV, $J$=$1.9$ eV) was adopted to reproduce the experimental band gap of LMO. However, such a large $U$ is abnormal for LMO according to literature \cite{Sawada:Prb56,Hashimoto:Prb,Lee:Prb13,Hou:Prb}. In fact, it is not physical meaningful to fit the experimental band gap by simply using overlarge $U$. Other methods like $GW$ calculation can be adopted to deal with the issue of band gap.

When the in-plane compressive strain (from STO substrate) is imposed on LMO bulk, i.e. by fixing the in-plane lattice constants of LMO to match the STO one and then relaxing the $c$-axis and atomic positions, the FM phase is indeed stabilized over the original A-AFM phases in the reasonable $U$ range, as shown by the orange curve in Fig.~\ref{fig2}(a) . With aforementioned default $U$(Mn) and $J$(Mn), the energy difference is $\sim26$ meV/Mn. Our conclusion of strain induced FM phase in LMO  qualitatively agrees with previous first-principles studies \cite{Nanda:Jmmm,Hou:Prb,Lee:Prb13}, although the concrete energy differences are different.

\subsection{III.B. LMO/STO heterostructures}
\begin{figure}
\includegraphics[width=0.48\textwidth]{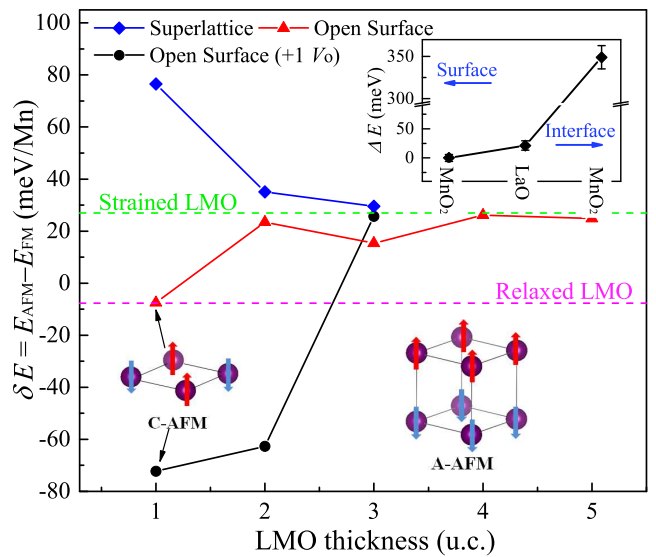}
\caption{Energy differences between two lowest magnetic states: in particular, C-type AFM and FM for $n=1$; A-AFM and FM for others. Other uniform magnetic orders such as C-type AFM are also calculated, which are much higher in energy. The energy differences between A-AFM and FM in unstrained and strained LMO bulks are also shown as dashed purple and green lines, respectively. Upper insert: the formation energy of one $V_{\rm O}$ as a function of layer positions in the $n=2$ open surface case. The formation energy of $V_{\rm O}$ in the outmost layer is taken as the reference. Lower insert: sketch of C-type AFM and A-AFM.}
\label{fig3}
\end{figure}

In this subsection, the LMO/STO heterostructures with various thicknesses will be calculated. The curves of energy are shown in Fig.~\ref{fig3}.

For heterostructure with open surface, when the LMO epitaxial film is only one u.c. thick (one LaO layer plus one MnO$_2$ layer), very interestingly, the magnetic ground state becomes C-type AFM, whose energy is lower than the FM state for $10$ meV/Mn. To our knowledge, it is first time to predict C-type AFM in LMO monolayer. When the LMO film is two u.c. thick, the ground state becomes FM, as in the strained bulk, although the energy difference between FM and AFM is a little lower than that of strained bulk. With further increasing thickness, as expected, the energy curve approach the strained bulk gradually. In short, only the monolayer limit of MnO$_2$ becomes magnetic ``dead" in our calculation, which does not fully agree with the experimental observation with the critical thickness of five u.c. \cite{Wang:Sci349}.

For comparison, the (LMO)$_n$/(STO)$_n$ ($n$=$1$, $2$, $3$) superlattices are also calculated, which shows clear FM tendency for all thickness. With increasing $n$, the energy curve also gradually approaches the strained LMO one. In fact, previous experiments on LSMO/STO superlattices indeed found that the strong FM magnetization could persist to ultra-thin limit (e.g. $4$ or $5$ u.c.) \cite{Kourkoutis:Pnas,Gray:Prb}, while the LSMO films with open surfaces are much easier to be ``dead" \cite{Huijben:Prb,Tebano:Prl}. And many technical details, e.g. size of laser spots during the pulse laser deposition, can affect the magnetization of LSMO layer \cite{Kourkoutis:Pnas}, implying the crucial role of non-stoichiometry.

\begin{figure}
\includegraphics[width=0.48\textwidth]{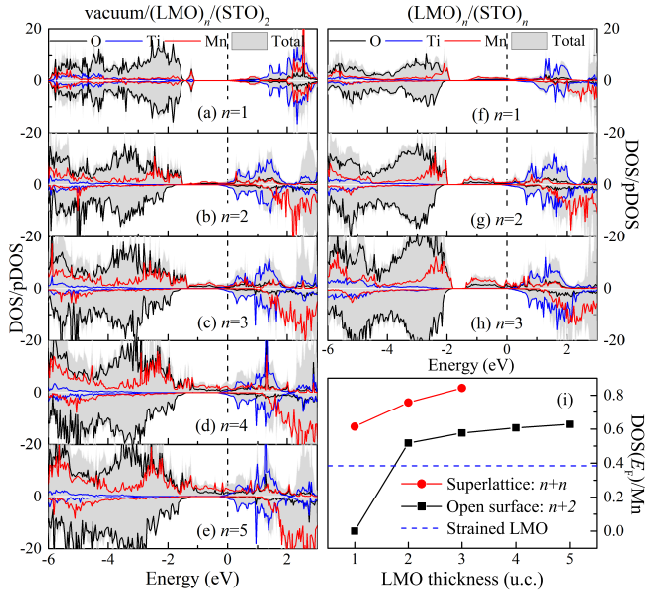}
\caption{Total DOS (grey shaded area) and pDOS (color lines) of LMO/STO heterostructures. (a-e) (LMO)$_n$/(STO)$_2$ heterostructures with open surfaces. (f-h) (LMO)$_n$/(STO)$_n$ superlattices. The vertical dashed line denotes the Fermi level. (i) The DOS value (per Mn) at the Fermi level as a function of LMO thickness.}
\label{fig4}
\end{figure}

To trace the evolution of electronic structure, the density of states (DOS) and atom-projected DOS (pDOS) of LMO/STO heterostructures are exhibited in Fig.~\ref{fig4}. In the series with open surface, (LMO)$_1$/(STO)$_2$ exhibits an insulating behavior with a gap of $\sim1.1$ eV. For thicker LMO, the heterostructures gradually become metallic, indicated by the enhanced DOS at Fermi level with increasing thickness (Fig~\ref{fig4}(i)). Although the experiment reported insulating behavior of ultra-thin LMO few layers \cite{Wang:Sci349}, the discrepancy may due to two reasons. First, as mentioned before, the band gaps are usually underestimated in DFT calculations. Second, the weak metallicity, i.e. small DOS values at Fermi level in our DFT calculations, can be easily suppressed in real materials due to intrinsic and extrinsic localizations. For comparison, the metallicity is much better in (LMO)$_n$/(STO)$_n$ superlattice, always with relative larger DOS values at Fermi level. The states of Ti are mostly above the Fermi level and thus empty. The charge transfer from Mn to Ti across the interface is evaluated by integrating the pDOS of Ti, which is only $\sim0.03$ electron per interfacial Ti (a negligible value), in consistent with previous reports \cite{Jilili:Sr,Liu:Jap}.

In experimental sample growth process of oxides, oxygen vacancies ($V_{\rm O}$'s) seem to be inevitable, more or less \cite{Siemons:Prl,Basletic:Nm,Aschauer:Prb,Cen:Nm,Zhong:Prb,Zhai:Nc}. Therefore, it is interesting to check the role of oxygen vacancies. Here one oxygen vacancy site is considered in our calculation. First, the forming energies of various $V_{\rm O}$'s are calculated, which suggest the topmost $V_{\rm O}$ is the most probable one (see the insert of Fig.~\ref{fig3}). By creating this $V_{\rm O}$, the system indeed becomes more likely to be antiferromagnetic. For the monolayer with one $V_{\rm O}$ ($16.7\%$), the C-AFM is lower in energy than the FM state for more than $70$ meV/Mn. And for the bilayer of LMO with one $V_{\rm O}$ ($8.3\%$), the A-AFM becomes the most stable one. Only starting from the trilayer of LMO with one $V_{\rm O}$ ($5.6\%$), the FM one takes over the ground state. It is rather complicated to consider multiple $V_{\rm O}$'s in thick LMO films due to too many possible combinations of $V_{\rm O}$ sites as well as combinations of possible non-uniform magnetic orders. Even though, our results for the single $V_{\rm O}$ cases already qualitatively indicate that the oxygen deficiency at surface is determinant for the disappearence of FM magnetization, or namely the magnetic ``dead layers".

\begin{figure}
\includegraphics[width=0.48\textwidth]{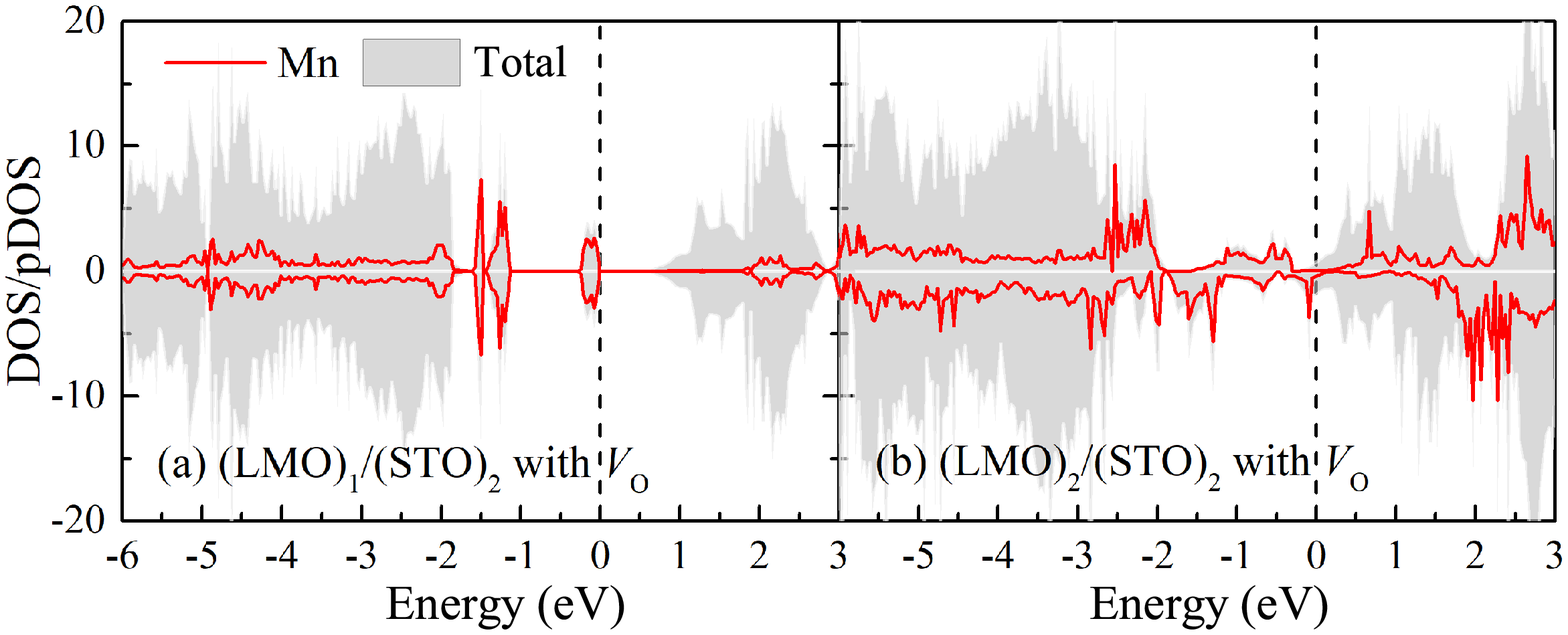}
\caption{The DOS and pDOS of Mn of non-stoichiometric (a) (LMO)$_1$/(STO)$_2$ and (b) (LMO)$_2$/(STO)$_2$ heterostructure, within which the surface oxygen is deficient.}
\label{fig5}
\end{figure}

The DOS and pDOS for heterostructures with one $V_{\rm O}$ are shown in Fig.~\ref{fig5}. The most significant change is the impurity state created near the Fermi level, which is quite reasonable.

In summary of our DFT results, the ferromagnetism of LMO is induced by the compressive strain from STO. The magnetic ``dead layers" in ultra-thin LMO film beyond monolayer are not intrinsic, but probably due to the oxygen vacancies near the surface.

\section{IV. Discussion}
\subsection{IV.A. Charge redistribution \& potential modulation}
The polar discontinuity has been well recognized as the origin of two-dimensional electron ``gas" (2DEG) in LaAlO$_3$/STO heterostructures \cite{Nakagawa:Nm,Son:Prb,Delugas:Prl}. In Ref.~\cite{Wang:Sci349}, the polar discontinuity of TiO$_2$ and LaO was argued to cause an electrostatic field, which led to charge redistribution. Wang \textit{et al.} proposed that the FM tendency beyond the critical thickness in LMO layers was due to this self-doping effect, while LMO itself should be non-FM \cite{Wang:Sci349}. This scenario is different from our DFT result mentioned before. Then it is necessary to check the polar discontinuity effect using DFT calculations.

\begin{figure}
\includegraphics[width=0.43\textwidth]{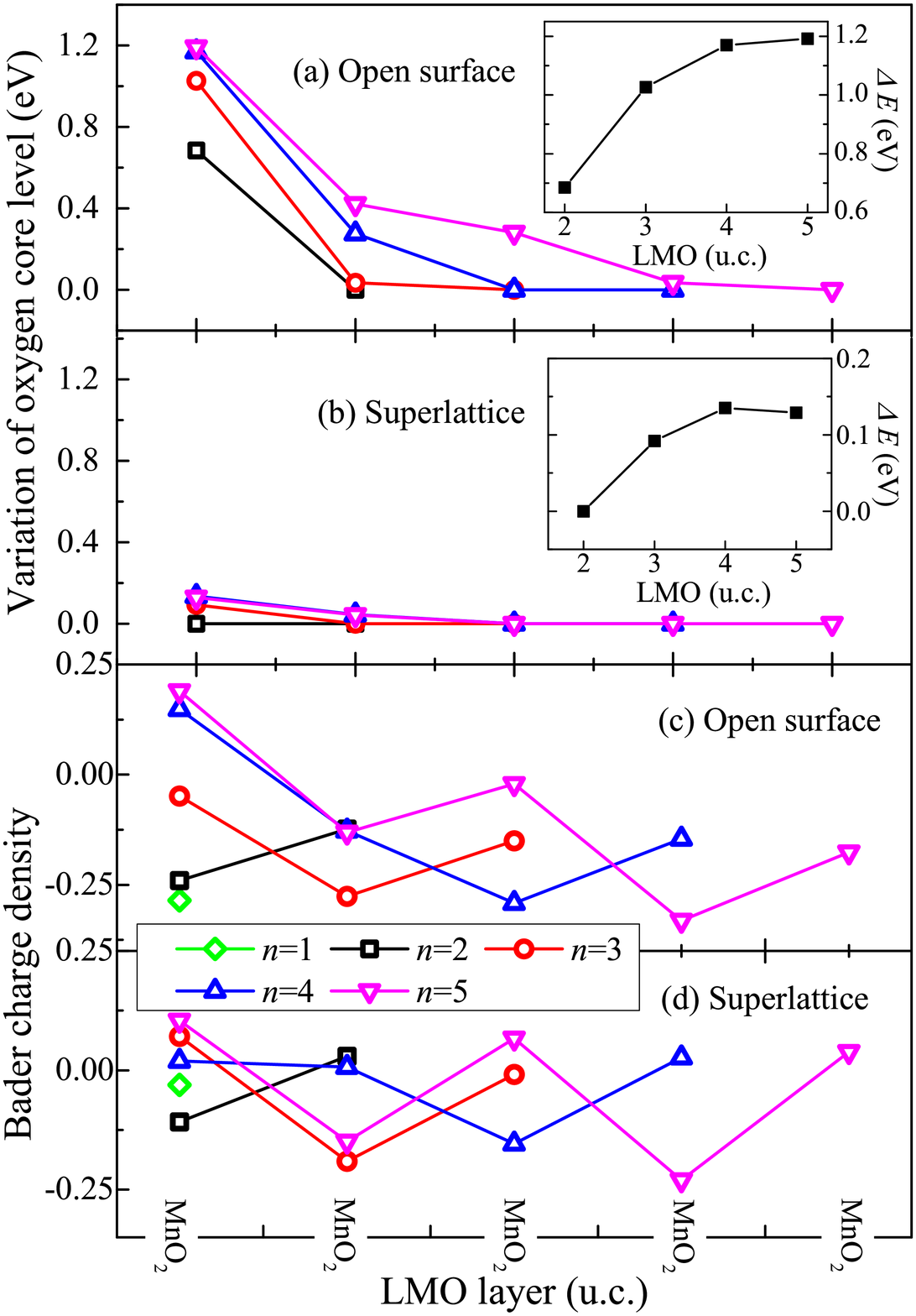}
\caption{Potential and electron density in LMO layers, counted from the surface or TiO$_2$-SrO-MnO$_2$ interface. (a-b) Energy profile of O's $1s$ orbital as a function of MnO$_2$ layer index, which is an indication of on-site potential from electrostatic field over the heterostructures. The O in interfacial MnO$_2$ layer is taken as the reference. (a) Superlattices. (b) Open surface cases. Inserts: The potential difference ($\Delta E$) between the top and bottom Mn's as a function of thickness. (c-d) Bader charge density of each MnO$_2$ layer. The reference density (value $0$) is set as the Bader charge density for MnO$_2$ layer in LMO bulk. Then a positive (negative) value means more electrons (holes) obtained. The oscillation of Bader charge density is very clear, while the unidirectional electron transfer from $p$-type interface (and surface) to $n$-type interface is obscure.}
\label{fig6}
\end{figure}

The on-site potential of MnO$_2$ layers can be estimated using oxygen $1s$ core level energy, as shown in Fig.~\ref{fig6}(a-b). First, the open surface cases show much larger modulation of potential, while the potential modulation in superlattice is rather mild. In fact, the polar discontinuity exists in both cases and should be similar in the magnitude before charge density redistribution. According to the calculated Bader charge for each MnO$_2$ layer (Fig.~\ref{fig6}(c-d)), there is indeed an overall tendency for electron transfer to the interface. If this electron transfer from $p$-type interface (or surface) to $n$-type interface is stronger, the electrostatic potential can be largely compensated. However, by comparing Fig.~\ref{fig6}(c) and Fig.~\ref{fig6}(d), the charge redistribution is more significant in the open surface cases. In this sense, the mild potential modulations in superlattices can not be attributed to the compensation by weak charge transfer.

Instead, the high dielectric constant of STO layers may compensate partial polar discontinuity in superlattices, as evidenced by the displacements of all Ti's along the same direction (see Fig.~\ref{fig6}(a) for more details).

For all heterostructures, no matter with or without open surfaces, there are clear charge density oscillation along the $c$-axis. This oscillation can be qualitatively understood as the effect from quantum confinement \cite{Dong:Prb13}. Such layer-dependent oscillation of electron density was once predicted in (LMO)$_n$/(LaNiO$_3$)$_n$ superlattices \cite{Dong:Prb13}, but rarely reported in the LaAlO$_3$/STO case. In LaAlO$_3$/STO case, the dominant driving force for charge redistribution is the polar field from termination, making the electron distribution behave in a decaying manner from the interface. In contrast, there¡¯s no polar discontinuity in LaNiO$_3$/LMO, and thus the dominant effect is the quantum confinement, which makes the electron distribution behave in an oscillation manner. Here the LMO/STO cases are in the middle of these two limits. The polar discontinuity exists but partially compensated by the many carriers in LMO layers. Noting that the maximum number of electron in (LaAlO$_3$)$_m$/(STO)$_n$ is one, distributing over $n$ layers of STO, while there are $n$ $e_{\rm g}$ electrons distributed in $n$ layers of LMO. Thus the polar discontinuity can be easily screened by one out of these $n$ electrons in the LMO/STO cases.

\begin{figure}
\includegraphics[width=0.48\textwidth]{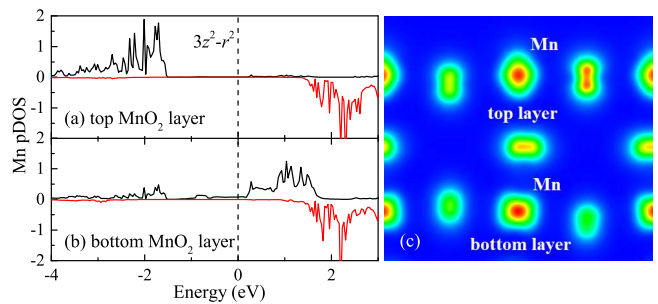}
\caption{Layer-resolved pDOS of Mn $3z^2-r^2$ orbital in (a) the top and (b) the bottom MnO$_2$ layer of vacuum/(LMO)$_2$/(STO)$_2$. (c) The electron density. The $3z^2-r^2$ character is obvious for the top layer.}
\label{fig7}
\end{figure}

In addition, the interaction between the interfacial STO layer (or vacuum) and LMO also tuned the charge density near interface (surface) a lot. Especially for the open surface, the oxygen octahedra are broken, which can seriously change the crystalline filed of Mn's $3d$ orbitals. Then the energy of $3z^2-r^2$ orbital for the surface Mn is greatly reduced, attracting more electrons to surface \cite{Calderon:Prb}. This conjecture can be further illustrated by the layer-resolved pDOS of Mn's $3z^2-r^2$ orbital, as shown in Fig.~\ref{fig7}.

In short, the potential modulation, as well as charge transfer in LMO layers, is driven by the collaborative effects of many issues, including polar discontinuity, dielectric screening of STO layer, quantum confinement, as well as interface interactions (or broken octahedra at surface), which are rather complicated. Then the simplified model of polar catastrophe established for the LaAlO$_3$/STO system \cite{Son:Prb,Delugas:Prl} can not be directly applied to explain the LMO/STO heterostructures.

\subsection{IV.B. Structural modulation}
\begin{figure}
\includegraphics[width=0.48\textwidth]{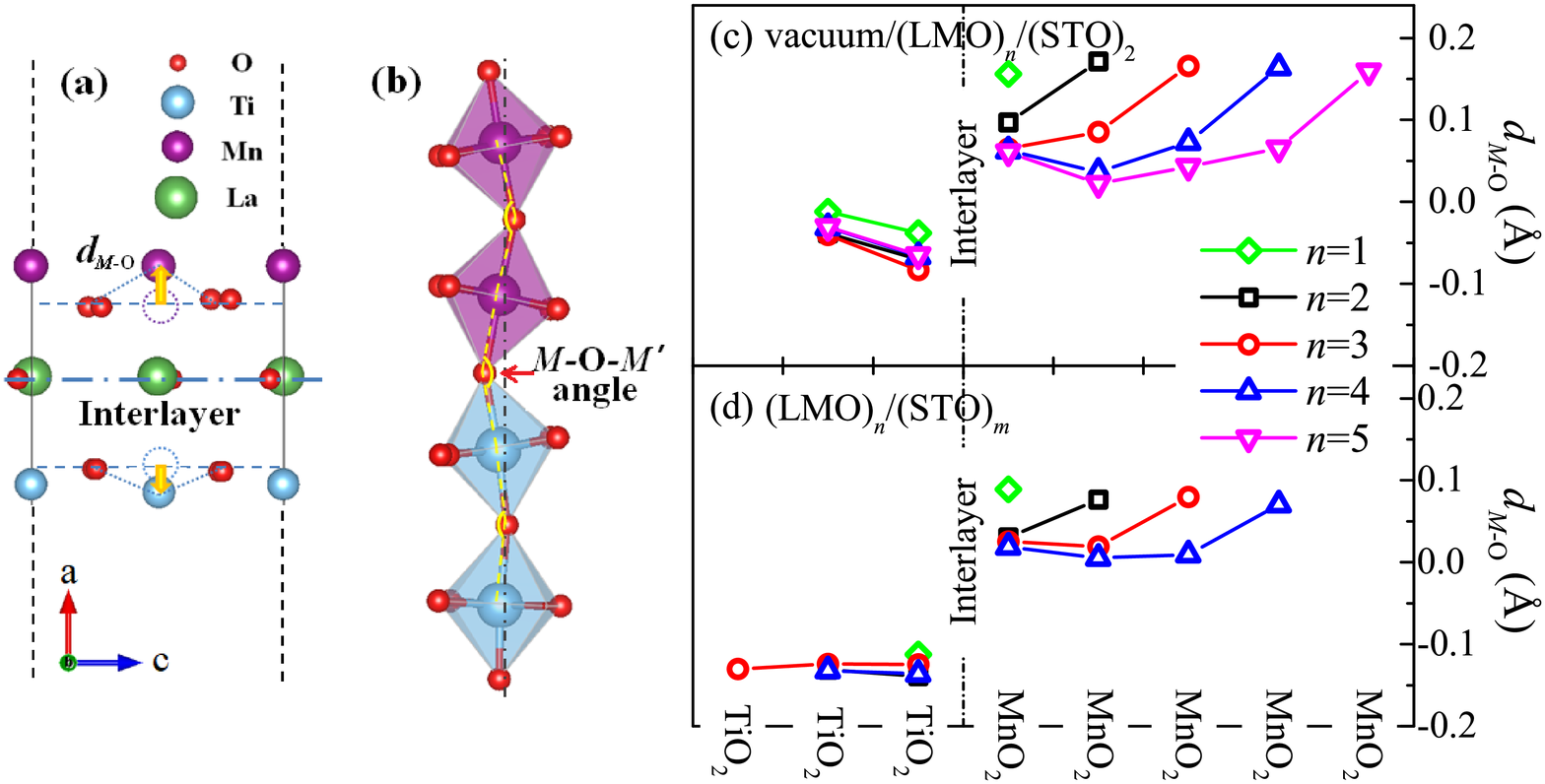}
\includegraphics[width=0.48\textwidth]{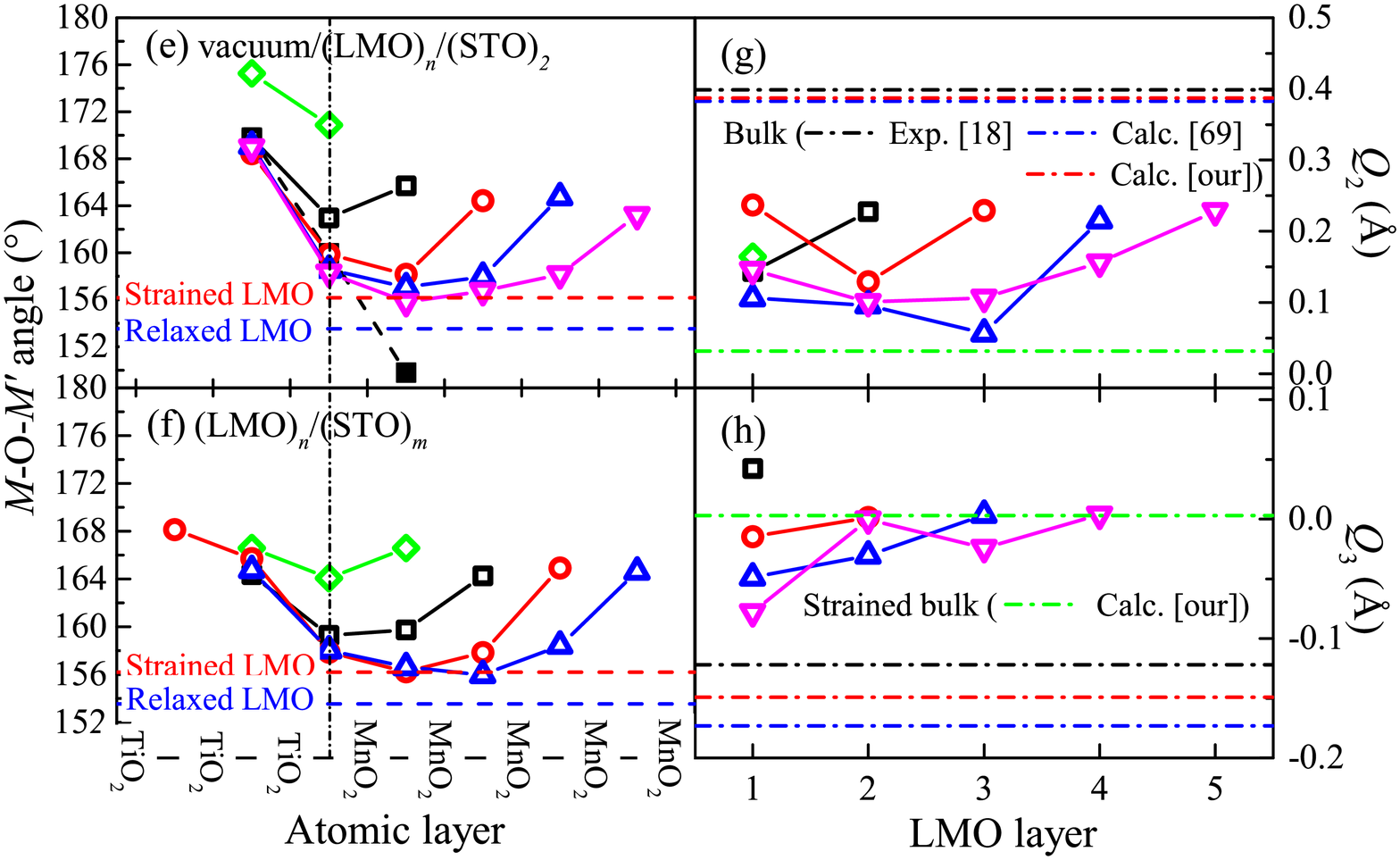}
\caption{Structural modulation in LMO/STO heterostructures. (a-b) Schematic of polar distortions and $M$O$_6$ octahedra tilting. (c-d) Polar distortions, characterized by the displacements of $M$, as a function of layer index. (e-f) $M$-O-$M'$ bond angles along the $c$-axis. Noting all end points in (f) are Mn-O-Ti bonds. The case with a $V_{\rm O}$ is also shown in (e) as solid black dots. (g-h) The JT distortions in the LMO portion of vacuum/(LMO)$_n$/(STO)$_2$ heterostructures. For comparison, both the experimental and theoretical values for LMO bulks are also presented \cite{Carvajal:Prb,Rivero:Prl}.}
\label{fig8}
\end{figure}

In oxide heterostructures and films, the magnetism is usually related to the structural modulations \cite{Sawada:Prb56,Hou:Prb,Lee:Prb13}. To further understand the LMO/STO heterostructures, the layer-resolved structural information is presented in Fig.~\ref{fig8}.

As stated in Sec.IV.A, the polar discontinuity of interface can be partially compensated by polar distortion of STO layer. In fact, both STO and LMO layers are distorted to be polarized, as shown in Fig.~\ref{fig8}(a). In all heterostructures, all Mn and Ti ions with positive charges move away from the TiO$_2$-LaO interface, as expected. Such polar distortions bend the in-plane O-Mn-O and O-Ti-O bond angles, which should be $180^\circ$ in corresponding bulks. Such polar ferromagnetism was recently experimentally reported in LSMO/BaTiO$_3$ superlattices \cite{Guo:Pnas}, which was attributed to be induced by ferroelectric distortion of BaTiO$_3$ \cite{Burton:Prb}. Here in our studied system, the polar discontinuity of interface can also induce polar ferromagnetism in ultra-thin LMO layers. The displacement of each ion (Mn or Ti) from the average height of corresponding O$_2$ is plotted in Fig.~\ref{fig8}(c-d). It is clearly that the open surface lead to more significant polar distortion due to the broken octahedra. At least for this polar structural modulation, the surface effect is stronger than the interfacial effect.

A recent experiment on LSMO/STO heterostructures revealed the layer dependent relaxation of MnO$_6$ octahedra tilting angles \cite{Liao:Nm}, which should have an important contribution to the magnetism. Generally, the tilting of octahedra bends Mn-O-Mn bonds, and thus changes the overlap between $3d$ and $2p$ orbitals, which tunes the exchange interactions. Here our DFT calculation can also provide the octahedra tilting evolution in ultra-thin LMO layers. Fig.~\ref{fig8}(e-f) shows the $M$-O-$M$' angles along the $c$ direction for optimized heterostructures. In pure LMO bulk, the Mn-O-Mn bond angle along the $c$-axis is $153^\circ$, which becomes straighter $155^\circ$ due to the compressive strain in the strained case.

In heterostructures, due to the non-tilting oxygen octahedra in STO layer, the tilting of octahedra in LMO layers is suppressed more or less, depending on the distance from the interface. The Mn-O-Ti bond angles at the TiO$_2$-LaO-MnO$_2$ interface are mostly around $158^\circ$ when LMO is beyond $2$ u.c.. While for the TiO$_2$-SrO-MnO$_2$ interface in superlattices, the Mn-O-Ti bond angles are around $165^\circ$. The inner Mn-O-Mn bond are usually more bending, closer to the intrinsic value of strained LMO. Even though, near the surface the Mn-O-Mn becomes straighter, which are all larger than $162^\circ$. However, once a $V_{\rm O}$ is created at surface, the out-of-plane Mn-O-Mn bond angle is significantly reduced to $\sim150^\circ$. Such $V_{\rm O}$ enhanced distortions can strongly suppress the FM tendency \cite{Tokura:Rpp,Dagotto:Prp}, in agreement with the result presented in Fig.~\ref{fig3}.

Besides the bending of bonds, the Jahn-Teller (JT) distortions also affect the properties of manganites \cite{Goodenough:Pr,Tokura:Rpp,Dagotto:Prp}. The evolutions of JT modes ($Q_2$ and $Q_3$) in heterostructures with open surfaces are calculated, as shown in Fig.~\ref{fig8}(g-h). For strained LMO bulk, the square geometry of STO (001) plane strongly suppresses the $Q_2$ mode, while the compressive strain strongly suppresses the $Q_3$ mode. For the LMO/STO heterostructures, the values of $Q_2$ and $Q_3$ are between the two limits of unstrained and fully strained cases, more closer to the latter. Thus the original $3x^2$-$r^2$/$3y^2$-$r^2$ orbital ordering should be generally suppressed, corresponding to the enhanced ferromagnetism. Especially, the $Q_2$ mode at the surface MnO$_2$ layer is largest.

In short, all these modulations of structural distortions suggest the crucial role of surface.

\section{V. Summary}
In the present study, with carefully verified parameters, our DFT calculations have investigated the physical properties of LMO and STO bulks, as well as strained LMO and LMO/STO superlattices. Our calculation indicates that FM state is the ground state for ideally strained LMO on STO. Thus, the appearance of ferromagnetism is intrinsic driven by compressive strain. The disappearance of such ferromagnetism in ultra-thin LMO few layers on STO is mostly due to open surface, which breaks the oxygen octahedra at surface and lead to structural and electronic reconstruction. Furthermore, the outmost oxygen vacancies are more likely to be created, which significantly suppress the ferromagnetism. Thus non-stoichiometry effect should play an important role in the experimentally observed non-magnetic state of ultra-thin LMO films.

According to our calculations, the polar discontinuity of interface can indeed induce charge redistribution near LMO/STO interface, but it is not the decisive effect for neither electron reconstruction nor disappearance of magnetization. The polar structural distortion can partially compensate the built-in electric field caused by polar discontinuity to a certain extent, making the self-doping effect weaker than expectation.

Although our study focused on LMO, the conclusion might be referential to understand and overcome the magnetic ``dead layers" widely existing in FM oxide films.

\acknowledgments{This study was supported by the National Natural Science Foundation of China (Grant Nos. 11674055 \& 11791240176). Most calculations were done on Tianhe-2 at National Supercomputer Centre in Guangzhou (NSCC-GZ).}

\bibliography{ref}
\end{document}